\RequirePackage{lineno}
\documentclass[prd,twocolumn,showpacs,amsmath,amssymb]{revtex4}
\usepackage{graphicx}
\usepackage{dcolumn}
\usepackage{bm}
\usepackage{psfig}
\usepackage{epsfig}
\usepackage{overpic}
\usepackage{verbatim}
\usepackage[colorlinks,linkcolor=blue,anchorcolor=red,citecolor=blue]{hyperref}

\usepackage{rotating}

\newcommand{\bfg}{\begin{figure}}
\newcommand{\efg}{\end{figure}}
\newcommand{\bitm}{\begin{itemize}}
\newcommand{\eitm}{\end{itemize}}
\newcommand{\bnum}{\begin{enumerate}}
\newcommand{\enum}{\end{enumerate}}
\newcommand{\btbl}{\begin{table}}
\newcommand{\etbl}{\end{table}}
\newcommand{\btbu}{\begin{tabular}}
\newcommand{\etbu}{\end{tabular}}

\newcommand{\beq}{\begin{equation}}
\newcommand{\edq}{\end{equation}}

\begin{document}
\normalsize
\parskip=5pt plus 1pt minus 1pt

\title{\boldmath Observation of a  structure in  $e^{+}e^{-} \to \phi \eta^{\prime}$ at $\sqrt{s}$ from 2.05 to 3.08 GeV}


\author{
M.~Ablikim$^{1}$, M.~N.~Achasov$^{10,d}$, P.~Adlarson$^{64}$, S. ~Ahmed$^{15}$, M.~Albrecht$^{4}$, A.~Amoroso$^{63A,63C}$, Q.~An$^{60,48}$, ~Anita$^{21}$, Y.~Bai$^{47}$, O.~Bakina$^{29}$
% ...
\\
(BESIII Collaboration)\\
$^{1}$ Institute of High Energy Physics, Beijing 100049, People's Republic of China\\
% ...
$^{4}$ Bochum Ruhr-University, D-44780 Bochum, Germany\\
% ...
$^{10}$ G.I. Budker Institute of Nuclear Physics SB RAS (BINP), Novosibirsk 630090, Russia\\
% ...
$^{15}$ Helmholtz Institute Mainz, Johann-Joachim-Becher-Weg 45, D-55099 Mainz, Germany\\
% ...
$^{21}$ Indian Institute of Technology Madras, Chennai 600036, India\\
% ...
$^{29}$ Joint Institute for Nuclear Research, 141980 Dubna, Moscow region, Russia\\
% ...
$^{47}$ Southeast University, Nanjing 211100, People's Republic of China\\
$^{48}$ State Key Laboratory of Particle Detection and Electronics, Beijing 100049, Hefei 230026, People's Republic of China\\
% ...
$^{60}$ University of Science and Technology of China, Hefei 230026, People's Republic of China\\
% ...
$^{63}$ (A)University of Turin, I-10125, Turin, Italy; (B)University of Eastern Piedmont, I-15121, Alessandria, Italy; (C)INFN, I-10125, Turin, Italy\\
$^{64}$ Uppsala University, Box 516, SE-75120 Uppsala, Sweden\\
% ...
$^{d}$ Also at the Novosibirsk State University, Novosibirsk, 630090, Russia\\
% ...
}


\date{\today}

\begin{abstract}

The process $e^{+}e^{-} \to \phi \eta^{\prime}$ has been studied for the first time in detail using data sample collected with the BESIII detector at the BEPCII collider at center of mass energies from 2.05 to 3.08 GeV. A resonance with quantum numbers $J^{PC}=1^{--}$ is observed with mass $M$ = (2177.5 $\pm$ 4.8 (stat) $\pm$ 19.5 (syst)) MeV/${ \it{c}^{\mathrm{2}}}$ and width $\Gamma$ = (149.0 $\pm$ 15.6 (stat) $\pm$ 8.9 (syst)) MeV with a statistical significance larger than 10$\sigma$, including systematic uncertainties. If the observed structure is identified with the $\phi(2170)$, then the ratio of partial width between the $\phi \eta^{\prime}$ by BESIII and $\phi \eta$ by {\it BABAR} is ($\mathcal{B}^{R}_{\phi \eta}\Gamma^{R}_{ee})/{(\mathcal{B}^{R}_{\phi \eta^{\prime}}\Gamma^{R}_{ee})}$ = 0.23 $\pm$ 0.10 (stat) $\pm$ 0.18 (syst), which is smaller than the prediction of the $s\bar{s}g$ hybrid models by several orders of magnitude.

\end{abstract}


\maketitle

\section{Introduction}

One of the most challenging questions in contemporary physics is how quarks and gluons form hadrons. Quantum chromodynamics (QCD) allows for any color-neutral combinations, however, a striking majority of all observed hadronic states are consistent with either a quark-antiquark pair, i.e., mesons, or triplet-quark systems, i.e., baryons. Although it has been difficult to unambiguously identify so-called exotic hadrons, such as glueballs, hybrids and multiquarks, remarkable progress has been made in the charm sector during the last decade. Some of those newly observed charmoniumlike or bottomoniumlike states are good candidates for exotics~\cite{ReviewOfModernPhysics, ReviewOfNaturePhysics, strangeoniumLike}. The strangeonium family may have states similar to those found in heavier quarkonia, and the more experimental information would be helpful to understand the prediction for the spectrum of strangeonium.

The $\phi(2170)$ was discovered in the process $e^{+}e^{-} \to \phi f_{0}(980)$ by {\it BABAR}~\cite{Y2175BABAR11, Y2175BABAR12} via the initial-state-radiation (ISR) technique, and was later confirmed by Belle~\cite{Y2175BELLE}, BES~\cite{Y2175BESII} and BESIII~\cite{Y2175BESIII, Y2175BESIII2019}. There are several interpretations of the $\phi(2170)$, including a regular $s\overline{s}$ meson in a $2^{3}D_{1}$~\cite{Y2175ss2D} or $3^{3}S_{1}$ configuration~\cite{Y2175SSbar2}, an $s \overline{s}g$ hybrid~\cite{Y2175hybrid, Y2175hybrid3}, a tetraquark state~\cite{Y2175tetraquark1,Y2175tetraquark2, Y2175tetraquark3, Y2175tetraquark4, Y2175tetraquark5}, a $\Lambda\overline{\Lambda}$ bound state~\cite{Y2175lambda,Y2175lambda1, Y2175lambda2, Y2175lambda3}, an $S$-wave threshold effect~\cite{SWaveThreshold}, or a three-meson system $\phi K^{+}K^{-}$~\cite{X2170}. The conventional $s\overline{s}$ meson is predicted to decay with significant fraction into the $s\bar{s}$-signature modes $\phi \eta$ and $\phi \eta^{\prime}$~\cite{Y2175SSbar2}. According to the Okubo-Zweig-Iizuka (OZI) rule~\cite{OZIRULE} and taking isospin effect into account, the contributions of $\omega$-like and $\rho$-like states are suppressed in the $\phi \eta$ and $\phi \eta^{\prime}$ modes. These two decay modes are useful to measure the mass and width of $\phi$-like states. On the other hand, the $s\bar{s}g$ hybrid state is expected to have a stronger coupling to $\phi \eta$, whose partial width is expected to be larger than that of $ \phi \eta^{\prime}$ by a factor of 3-200~\cite{Y2175hybrid, Y2175hybrid3}. The ratio between the $\phi \eta $ and $\phi \eta^{\prime}$ decay widths is therefore an important observable to test $\phi(2170)$ as a hybrid state.

BESIII measured the processes $e^{+}e^{-} \to K^{+}K^{-}$~\cite{ResoanceBESIII2019} and $e^{+}e^{-} \to \Lambda \bar{\Lambda}$~\cite{LambdaAntiLambda} to test the prediction of $\phi(2170)$ as the $\Lambda \bar{\Lambda}$ bound state. An enhancement at $\sqrt{s}$ = 2.232 GeV in the process $e^{+}e^{-} \to \phi K^{+}K^{-}$~\cite{ResoanceBESIII2019PhiKK} is difficult to be interpreted by the Faddeev calculation for the three-meson system $\phi K \bar{K}$. Assuming that the observed structure in the process $e^{+}e^{-} \to K^{+}(1460)K^{-}$ is $\phi(2170)$, it implies that the theoretical expectation for the hybrid state is not in agreement with the experimental results~\cite{ResoanceBESIII2020}. {\it BABAR} observed evidence of $\phi(2170)$ in a study of the process $e^{+}e^{-} \to \gamma_{\mathrm{ISR}} \phi \eta$~\cite{PRDPhiEta2008} and a small signal in $e^{+}e^{-} \to  \gamma_{\mathrm{ISR}} \phi \eta^{\prime}$~\cite{PRD76092005}. The tail of the $\phi(1680)$ contributes to the $\phi \eta$ mode. However, $\phi(1680)$ decays into $\phi \eta^{\prime}$ are highly suppressed~\cite{Phi1680}. In a BESIII study of $J/\psi \to \eta \phi \eta^{\prime}$, evidence was found of a structure in the $\phi \eta^{\prime}$ mass spectrum in the $2.0 - 2.1$ GeV/$c^{2}$ region under the assumption of $J^{P} = 1^{-}$~\cite{BESIIIPhiEtaEtap}. The $e^{+}e^{-} \to \phi \eta^{\prime}$ process provides therefore important input for the understanding of the $\phi(2170)$.

In this paper, we present a measurement of the Born cross sections of $e^{+}e^{-} \to \phi \eta^{\prime}$ as a function of center-of-mass (c.m.) energies from 2.05 to 3.08 GeV based on 20 data samples corresponding to an integrated luminosity of 640 pb$^{-1}$ collected at the Beijing spectrometer (BESIII).

\section{DETECTOR AND DATA SAMPLES}

The BESIII detector is a magnetic spectrometer~\cite{BESIII} located at the Beijing Electron Positron Collider (BEPCII)~\cite{BEPCII}. The cylindrical core of the BESIII detector consists of a helium-based multilayer drift chamber (MDC), a plastic scintillator time-of-flight system (TOF), and a CsI(Tl) electromagnetic calorimeter (EMC), which are all enclosed in a superconducting solenoidal magnet providing a 1.0~T magnetic field. The solenoid is supported by an octagonal flux-return yoke with resistive plate counter muon identifier modules interleaved with steel. The acceptance of
charged particles and photons is 93\% over $4\pi$ solid angle. The charged-particle momentum resolution at $1~{\rm GeV}/c$ is $0.5\%$, and the $dE/dx$ resolution is $6\%$ for the electrons
from Bhabha scattering. The EMC measures photon energies with a resolution of $2.5\%$ ($5\%$) at $1$~GeV in the barrel (end cap) region. The time resolution of the TOF barrel part is 68~ps, while that of the end cap part is 110~ps.

A {\sc geant4}-based~\cite{Geant4} Monte Carlo (MC) simulation software is used to generate simulated data samples. The software implementation includes geometric and material description of the BESIII detector, the detector response and digitization models. It also accounts for the variation in detector running conditions and performance.

To study backgrounds, a generic MC sample for the process $e^{+}e^{-} \to q \bar{q}$ with $q = u, d, s$ is generated  with  {\sc conexc}~\cite{ConExc}, while the hadronization processes are generated by {\sc evtgen}~\cite{EVENTGEN1, BESEVENTGEN2} for known modes with branching fractions set to Particle Data Group (PDG) world average values~\cite{PDG} and by {\sc luarlw}~\cite{LUARLW} for the remaining unknown decays. The signal MC sample for $e^{+}e^{-} \to \phi \eta^{\prime}$ is also generated by {\sc conexc}, taking radiative corrections, the angular distributions of the final state and the amplitude of $\eta^{\prime} \to \gamma \pi^{+} \pi^{-}$~\cite{AmplitudeOfEtap} into account at each c.m. energy point.

\section{EVENT SELECTION AND BACKGROUND ANALYSIS}

The MC simulations are used to optimize the selection criteria, the determination of detection efficiencies and estimation of the background. Taking the branching fractions of the decays of intermediate states and the efficiency of photon detection into consideration for the process of $e^{+}e^{-} \to \phi \eta^{\prime}$, the $\phi$ candidate is identified from a $K^{+}K^{-}$ pair and the $\eta^{\prime}$ from $\pi^{+}\pi^{-}\gamma$ combinations. 
To improve the detection efficiency, candidate events are required to have three or four good charged tracks, corresponding to two detected pions and one or two detected kaons, and at least one good photon. Tracks are reconstructed from hits in the multilayer drift chamber (MDC) within $|\cos\theta|<0.93$, where $\theta$ is the polar angle with respect to the magnetic field direction. The tracks are required to pass the interaction point within 10~cm along the beam direction and within 1~cm in the transverse direction to the beam. For each charged track, the time-of-flight (TOF) from scintillation counters and the energy loss measurement ($dE/dx$) information from MDC are combined to form particle identification confidence levels (C.L.) for the $\pi$, $K$, and $p$ hypotheses. The particle type with the highest C.L.\ is assigned to each track. Two pions and at least one kaon are required per event. 

Photon candidates are selected from showers in the electromagnetic calorimeter (EMC) that are not associated with charged tracks. Good photon candidates reconstructed in the barrel part of the EMC must have a polar angle within $|\cos\theta|<0.8$ and a minimum deposited energy of 25 MeV. To be reconstructed in the end caps, the photon candidates must have a polar angle within $0.86<|\cos\theta|< 0.92$  and a minimum energy deposit of 50 MeV. Timing information in the EMC is used to suppress electronic noise and energy deposits unrelated to the event.
In order to suppress the background from ISR processes, the energy of photon candidates is required to be larger than 70 MeV. 
The tracks and photon candidates are then combined and subject to further analysis. 
The interaction vertex of the event is reconstructed by two pions and one kaon. 
A one-constraint (1C) kinematic fit is performed under the hypothesis that the $K\pi^{+}\pi^{-}\gamma$ missing mass corresponds to the kaon mass~\cite{PDG}. If both kaons are identified in an event, the combination with the smallest $\chi^{2}$ of the 1C kinematic fit is retained. The corresponding $\chi^{2}$, denoted as $\chi^{2}_{1C}(\pi^{+}\pi^{-}KK_{\mathrm{miss}}\gamma)$, is required to be smaller than 20. The candidate event of $e^{+}e^{-} \to \phi \eta^{\prime}$ is required to be within the $\phi$ signal region, defined as $|M(KK_{\mathrm{miss}}) - m_{\phi}|<3\sigma$, where $m_{\phi}$ is the nominal $\phi$ mass from PDG and $\sigma$ is the $\phi$ width convolved with detector resolution. The sideband region, defined as 1.050 GeV/$c^{2}<M(KK_{\mathrm{miss}})<$1.130 GeV/$c^{2}$, is used to estimated the non-$\phi$ background contributions. 

A study of the $e^{+}e^{-} \to q \bar{q}$ MC sample shows that the dominant background processes are $e^{+}e^{-} \to \phi \pi^{+}\pi^{-}$, $e^{+}e^{-} \to K K^{*} \pi $ and $e^{+}e^{-} \to K^{+}K^{-} \rho(770)$.  No peaking background is observed in the signal region of the $\pi^{+}\pi^{-}\gamma$ invariant-mass distribution. 

 \section{DETERMINATION OF THE BORN CROSS SECTION}
\begin{figure}[htbp]
\begin{center}
\begin{overpic}[width=8.5cm,height=6.5cm,angle=0]{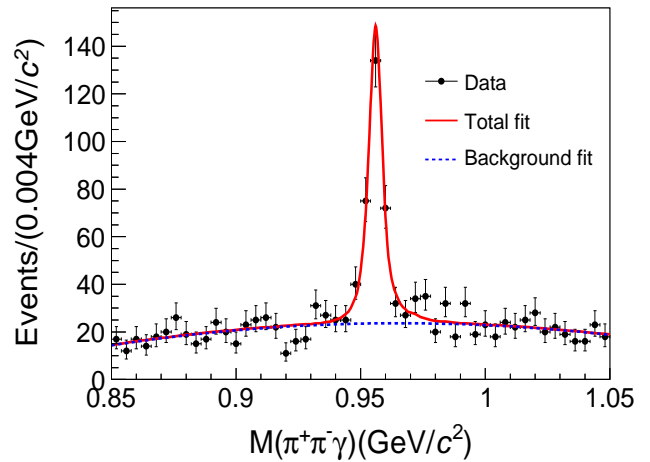}
\end{overpic}
\end{center}
\caption{Fit to the $M(\pi^{+}\pi^{-}\gamma)$ mass spectrum at $\sqrt{s}$ = 2.1250 GeV. The black dots with error bars are  data, the solid (red) curve is the total fit result and the dashed (blue) curve is the background shape.}
\label{Fitting2125}
\end{figure}

The $e^{+}e^{-} \to \phi \eta^{\prime}$ signal yield is determined by performing an unbinned maximum likelihood fit to the $\pi^{+}\pi^{-}\gamma$ invariant-mass distribution. The signal is described by the line-shape obtained from the signal MC simulation convolved with a Gaussian function that accounts for the difference in resolution between data and MC simulation. The shape of the background is parametrized by a second-order polynomial function. The corresponding fit result is shown in Fig.~\ref{Fitting2125} at $\sqrt{s}= 2.1250$ GeV.

The same event selection criteria and fit procedure are applied to the other 19 data samples taken at different c.m.~energies. The numbers of signal events for these samples are listed in Table~\ref{XSectionSummaryPhiEtapEnergyPoints11}.


The Born cross section is calculated using
\begin{center}
\begin{equation}
\sigma^{B}=\frac{N^{\mathrm{obs}}}{\mathcal{L}\cdot (1+\delta)
\cdot \epsilon \cdot \cal {B}} ,
\label{EquationPhiEtap}
\end{equation}
\end{center}

\noindent
where $\mathit{N^{\mathrm{obs}}}$ is the number of signal events, $\mathcal{L}$ the integrated luminosity measured with the method described in Ref.~\cite{LuminosityFinal}, $\cal B$ the product of the branching fractions of the decays $\phi \to K^{+} K^{-}$ and $\eta^{\prime} \to \pi^{+} \pi^{-} \gamma$~\cite{PDG}, $\epsilon$ the detection efficiency and $(1+\mathrm{\delta})$  is the correction factor due to ISR and vacuum polarization (VP). Both $\epsilon$ and $(1+\mathrm{\delta})$  are obtained from MC simulations of the signal reaction at the individual c.m.~energies~\cite{ISR, VP}. The detection efficiency and ISR factor depend on the input Born cross section, where the iterations are performed until the measured Born cross section does not change by more than 1.0$\%$. The resulting Born cross sections and related variables are listed in Table~\ref{XSectionSummaryPhiEtapEnergyPoints11}.


\begin{table}[htbp]
  \centering
  \caption{The Born cross sections of $e^{+}e^{-} \to \phi \eta^{\prime}$. The center-of-mass energy ($\sqrt{s}$), integrated luminosity ($\mathcal{L}$), the yields of signal events ($\mathit{N^{\mathrm{obs}}}$),  the product of radiative correction factor and vacuum polarization factor $(1+\mathrm{\delta})$, detection efficiency ($\mathrm{\epsilon}$), Born cross section ($\sigma^{\mathit{B}}$). The first uncertainties are statistical and the second systematic.}
   \newsavebox{\tablebox}
   \begin{lrbox}{\tablebox}
  
  \begin{tabular}{l c  c c c c}
  \hline
  \hline
  $\sqrt{s}$ (GeV) &$\cal{L}$ (pb$^{-1}$)      &$\mathit{N^{\mathrm{obs}}}$            &$(1+\delta)$                              &$\epsilon$         &$\sigma^{\mathit{B}}$ (pb)\\
  \hline
2.0500		&3.34		&4.3$\pm$3.0		&0.888		&0.257		&39.7$\pm$27.7$\pm$2.4\\
2.1000		&12.2		&21.3$\pm$6.3		&0.926		&0.290		&45.9$\pm$13.6$\pm$2.8\\
2.1250		&108		&267.7$\pm$22.2		&0.938		&0.299		&61.8$\pm$5.1$\pm$3.6\\
2.1500		&2.84		&12.3$\pm$4.2		&0.948		&0.310		&103.6$\pm$35.4$\pm$6.0\\
2.1750		&10.6		&87.4$\pm$11.0		&0.957		&0.324		&186.5$\pm$23.5$\pm$11.4\\
2.2000		&13.7		&105.5$\pm$11.8		&0.964		&0.327		&171.9$\pm$19.2$\pm$10.3\\
2.2324		&11.9		&73.6$\pm$10.2		&0.972		&0.331		&135.6$\pm$18.8$\pm$8.4\\
2.3094		&21.1		&65.6$\pm$9.8		&0.977		&0.339		&66.1$\pm$9.9$\pm$4.2\\
2.3864		&22.5		&52.7$\pm$8.8		&0.992		&0.341		&48.6$\pm$8.1$\pm$3.7\\
2.3960		&66.9		&163.9$\pm$15.0		&0.994		&0.343		&50.6$\pm$4.6$\pm$2.9\\
2.5000		&1.10		&3.7$\pm$2.1		&1.007		&0.347		&67.8$\pm$38.5$\pm$4.3\\
2.6444		&33.7		&73.9$\pm$9.5		&1.009		&0.348		&43.9$\pm$5.6$\pm$2.3\\
2.6464		&34.0		&50.4$\pm$8.0		&1.009		&0.346		&29.8$\pm$4.7$\pm$1.6\\
2.8000		&1.01		&2.0$\pm$1.4		&0.996		&0.352		&39.8$\pm$27.9$\pm$2.0\\
2.9000		&105		&113.3$\pm$12.0		&1.011		&0.346		&21.6$\pm$2.3$\pm$1.2\\
2.9500		&15.9		&9.9$\pm$3.4		&1.013		&0.343		&12.6$\pm$4.3$\pm$0.7\\
2.9810		&16.1		&6.9$\pm$2.9		&1.012		&0.343		&8.7$\pm$3.7$\pm$0.7\\
3.0000		&15.9		&12.6$\pm$3.7		&1.011		&0.342		&16.2$\pm$4.7$\pm$0.8\\
3.0200		&17.3		&14.5$\pm$4.1		&1.008		&0.340		&17.2$\pm$4.9$\pm$0.9\\
3.0800		&126		&90.2$\pm$10.3		&0.906		&0.339		&16.4$\pm$1.9$\pm$1.0\\
\hline
    \end{tabular}
    
    \end{lrbox}
    \scalebox{0.85}{\usebox{\tablebox}}

    \label{XSectionSummaryPhiEtapEnergyPoints11}
\end{table}

\section{SYSTEMATIC UNCERTAINTY}

The following sources of systematic uncertainties are considered in the measurement of the Born cross sections. The common uncertainties include the integrated luminosity, the tracking efficiency, photon detection, PID and branching fractions of intermediate state decays for each energy point. The systematic uncertainties also arise from the kinematic fit, the fit procedure, mass window requirement of $\phi$, ISR correction factor, as well as MC statistics. The uncertainty of the integrated luminosity is $1\%$ at each energy point~\cite{LuminosityFinal}. The uncertainty of the efficiency for each charged track and PID are estimated to be 1$\%$~\cite{ResoanceBESIII2019}. The uncertainty due to photon detection is 1$\%$~\cite{Photon}. The uncertainty of the branching fractions of intermediate states is taken from the PDG~\cite{PDG}, it is 2.2 $\%$.  The uncertainty related to the kinematic fit is estimated by correcting the helix parameters of the simulated charged tracks to match the resolution in data~\cite{Aixiaocong}. The difference in $(1+\delta)\epsilon$ between the last two iterations of the cross section measurement is taken as the uncertainty related to the ISR correction factor. The $\phi$ line shape in simulated data is smeared to better match the data line shape. The difference in the detection efficiency before and after smearing are assigned as systematic uncertainties for the $\phi$ mass window requirement. The difference in the signal yield between fits in a range of (0.8, 1.10) GeV/$c^{2}$ compared to the nominal fit is treated as the systematic uncertainty from the fit range. The uncertainty related to the signal shape is estimated with an alternative fit using the same function for the signal shape, but fixing the width of the Gaussian function to the value obtained in the nominal fit plus one standard deviation. The background shape is described as a second-order polynomial function. A fit with a third-order polynomial function for the background shape is used to estimate the uncertainty. The uncertainty due to MC statistics is estimated by the number of the generated events. Assuming that all of the above systematic uncertainties are uncorrelated, the total systematic uncertainties are obtained by adding the individual uncertainties in quadrature, shown in Table~\ref{XTotalSystematicPhiEtap}.


\begin{table*}[htbp]
  \centering
  \caption{Relative systematic uncertainties (in $\%$) in the Born cross sections of $e^{+}e^{-} \to \phi \eta^{\prime}$. These represent the uncertainties in the estimated effects of the luminosity ($\mathcal{L}$), tracking efficiency (Tracking), photon reconstruction efficiency (Photon), PID efficiency (PID), the kinematic fit (KinFit), signal and background shape (Signal and Background), fit range (Range), the initial state radiation factor (ISR), $\phi$ mass window ($m_{\phi}$), MC statistics (MC), and branching fraction ($\mathcal{B}$). The total uncertainty is obtained by summing the individual contributions in quadrature.}
  
\vskip5pt
  
\begin{lrbox}{\tablebox}
\begin{tabular}{cccccccccccccccccccccc}
\hline
\hline
$\sqrt{s}$ (GeV)    &$\mathcal{L}$   &Tracking  &Photon &PID   &KinFit      &Signal     &Background     &Range  &ISR &&   &$m_{\phi}$   &&    &MC  &&  &$\mathcal{B}$   &&      & Sum\\
\hline
2.0500  &1.0  &3.0  &1.0  &3.0  &3.0  &0.0  &0.0  &1.5  &0.1  &&  &0.5  &&  &0.5  &&  &2.2  &&  &6.0\\
2.1000  &1.0  &3.0  &1.0  &3.0  &2.7  &0.0  &0.9  &1.5  &0.9  &&  &0.5  &&  &0.5  &&  &2.2  &&  &6.0\\
2.1250  &1.0  &3.0  &1.0  &3.0  &2.6  &0.3  &0.4  &1.5  &0.8  &&  &0.5  &&  &0.5  &&  &2.2  &&  &5.9\\
2.1500  &1.0  &3.0  &1.0  &3.0  &2.3  &0.0  &0.8  &1.2  &0.8  &&  &0.5  &&  &0.5  &&  &2.2  &&  &5.8\\
2.1750  &1.0  &3.0  &1.0  &3.0  &2.0  &1.5  &0.7  &2.1  &1.0  &&  &0.4  &&  &0.5  &&  &2.2  &&  &6.1\\
2.2000  &1.0  &3.0  &1.0  &3.0  &2.3  &0.8  &1.6  &1.3  &0.7  &&  &0.5  &&  &0.5  &&  &2.2  &&  &6.0\\
2.2324  &1.0  &3.0  &1.0  &3.0  &2.4  &1.9  &1.5  &1.1  &0.1  &&  &0.4  &&  &0.4  &&  &2.2  &&  &6.2\\
2.3094  &1.0  &3.0  &1.0  &3.0  &1.9  &2.0  &2.9  &0.6  &0.2  &&  &0.4  &&  &0.4  &&  &2.2  &&  &6.4\\
2.3864  &1.0  &3.0  &1.0  &3.0  &1.9  &2.5  &4.7  &1.4  &0.3  &&  &0.4  &&  &0.4  &&  &2.2  &&  &7.7\\
2.3960  &1.0  &3.0  &1.0  &3.0  &1.6  &0.1  &2.0  &0.9  &0.6  &&  &0.4  &&  &0.4  &&  &2.2  &&  &5.7\\
2.5000  &1.0  &3.0  &1.0  &3.0  &1.6  &0.0  &2.7  &2.4  &0.9  &&  &0.4  &&  &0.4  &&  &2.2  &&  &6.4\\
2.6444  &1.0  &3.0  &1.0  &3.0  &0.8  &0.0  &0.8  &0.9  &0.0  &&  &0.4  &&  &0.4  &&  &2.2  &&  &5.2\\
2.6464  &1.0  &3.0  &1.0  &3.0  &0.8  &0.0  &1.4  &1.6  &0.2  &&  &0.4  &&  &0.4  &&  &2.2  &&  &5.5\\
2.8000  &1.0  &3.0  &1.0  &3.0  &0.5  &0.0  &0.0  &0.2  &0.4  &&  &0.4  &&  &0.4  &&  &2.2  &&  &5.1\\
2.9000  &1.0  &3.0  &1.0  &3.0  &0.3  &1.3  &0.3  &1.6  &0.1  &&  &0.4  &&  &0.4  &&  &2.2  &&  &5.4\\
2.9500  &1.0  &3.0  &1.0  &3.0  &0.1  &1.0  &0.0  &2.9  &0.3  &&  &0.4  &&  &0.4  &&  &2.2  &&  &5.9\\
2.9810  &1.0  &3.0  &1.0  &3.0  &0.0  &0.0  &0.0  &5.6  &0.2  &&  &0.4  &&  &0.4  &&  &2.2  &&  &7.5\\
3.0000  &1.0  &3.0  &1.0  &3.0  &0.1  &0.0  &0.8  &0.6  &0.6  &&  &0.4  &&  &0.4  &&  &2.2  &&  &5.1\\
3.0200  &1.0  &3.0  &1.0  &3.0  &0.0  &0.7  &2.1  &0.5  &0.2  &&  &0.4  &&  &0.4  &&  &2.2  &&  &5.5\\
3.0800  &1.0  &3.0  &1.0  &3.0  &0.1  &0.9  &2.0  &1.7  &1.0  &&  &0.5  &&  &0.4  &&  &2.2  &&  &5.8\\

\hline
\hline
\end{tabular}   
\end{lrbox}
\scalebox{1.1}{\usebox{\tablebox}}   
\label{XTotalSystematicPhiEtap}
\end{table*}

\section{Fit to the line shape}

\begin{figure}[htbp]
\begin{center}
\begin{overpic}[width=7.5cm,height=5.5cm,angle=0]{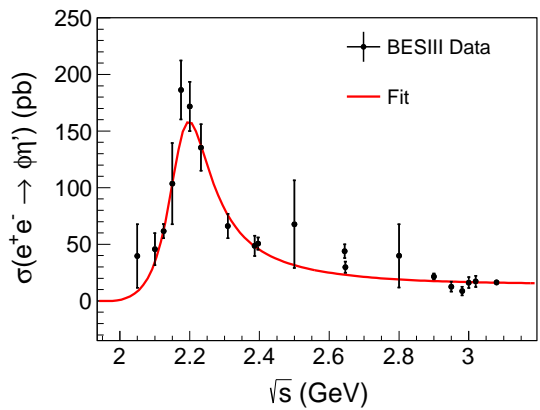}
\end{overpic}
\end{center}
\caption{Born cross sections of the $e^{+}e^{-} \to \phi \eta^{\prime}$ process. The solid curve (red) shows the fit to the line shape of the Born cross sections. The dots (black) with error bars show data.  }
\label{XSectionComparisonSummary}
\end{figure}

The measured Born cross sections are shown in Fig.~\ref{XSectionComparisonSummary},  where a clear structure is observed around 2.2 GeV. To study a possible resonant behavior, a $\chi^{2}$ fit incorporating the correlated and uncorrelated uncertainties is performed to the measured Born cross sections. Assuming that the final states $ \phi \eta^{\prime}$  come from a resonance decay, we fit the line shape using a coherent sum of a phase-space modified Breit-Wigner (BW) function and a phase-space term. The probability density function (PDF) is defined as

\begin{center}
\begin{equation}
\left| A(\sqrt{s})\mid^{2}=\mid C_{0}\sqrt{\Phi(\sqrt{s})}+e^{i\varphi} \times BW(\sqrt{s}) \right|^{2},
\label{FittingLineshapeOfCrossSection1}
\end{equation}
\end{center}

\noindent
where the BW function is written as

\begin{center}
\begin{equation}
BW(\sqrt{s})=\frac{M_{R}}{\sqrt{s}}\frac{\sqrt{12\pi\Gamma_{e^{+}e^{-}}^{R}\mathcal{B_{R}(\phi \eta^{\prime})}\Gamma_{\text{tot}}^{R}}}{s-M_{R}^{2}+iM_{R}\Gamma_{\text{tot}}^{R}} \cdot \sqrt{\frac{\Phi(\sqrt{s})}{\Phi(M_{R})}},
\label{FittingLineshapeOfCrossSection2}
\end{equation}
\end{center}

\noindent
where $M_{R}$ is the mass of the resonance, $\Gamma_{\text{tot}}^{R}$ the total width, $\Gamma_{e^{+}e^{-}}^{R}$ the $e^{+}e^{-}$ partial width, $\mathcal{B_{R}(\phi \eta^{\prime})}$ the branching fraction of the resonance decay to $\phi \eta^{\prime}$, $\varphi$ the phase angle between the resonance and the phase-space contribution and $\Phi(\sqrt{s})$ the phase space factor for a $P$-$\text{wave}$ two-body system.

The fit has two solutions with an identical mass and width of the resonance. The product $\Gamma_{e^{+}e^{-}}^{R} \mathcal{B_{R}(\phi \eta^{\prime})}$  is also the same in the two solutions, while the phases are different.  
The fit quality is estimated by inspecting the $\chi^{2}$, which gives a $\chi^{2}/\mathrm{ndf} = 28.30/15$, where ndf is the number of degrees of freedom. 
The parameters of the structure are determined to be $M = (2177.5 \pm 4.8)$ MeV/$c^{2}$ and $\Gamma = (149.0 \pm 15.6)$ MeV, where the uncertainty is statistical only. Figure~\ref{XSectionComparisonSummary} shows the fit result, and the parameters of the resonance are summarized in Table~\ref{ParametersFromResonance}. The significance of the resonance is determined to be larger than 10$\sigma$, including systematic uncertainties. This is obtained by comparing the change of $\Delta(\chi^{2})$ with and without the resonance in the fit and taking the change in the number of degrees of freedom $\Delta \text{ndf}$ = 4 into account. 


\begin{table}[htbp]
  \centering
  \caption{Line shape parameters obtained by the fit. }
  \begin{tabular}{c|c}
  \hline
  \hline
  Parameter                                    	&Solution $\uppercase\expandafter{\romannumeral1}$  \qquad  	Solution $\uppercase\expandafter{\romannumeral2}$\\
  \hline
  $M_{R}$ (MeV/$c^{2}$)         		&2177.5 $\pm$ 4.8 (stat) $\pm$	19.5 (syst)		   \\

   $\Gamma_{\text{tot}}^{R}$ (MeV)			&149.0 $\pm$ 15.6 (stat)$\pm$	8.9 (syst)	   \\

  $\mathcal{B_{R} }\Gamma_{e^{+}e^{-}}^{R}$(eV)      	                 &7.1$\pm$0.7 (stat) $\pm$0.7 (syst)       \\

  $\varphi$(rad)		&3.13	$\pm$ 2.01	\qquad		-0.01 $\pm$ 2.36\\
    \hline
  \hline
    \end{tabular}
    \label{ParametersFromResonance}
\end{table}


\begin{figure}[htbp]
\begin{center}
\begin{overpic}[width=7.5cm,height=5.5cm,angle=0]{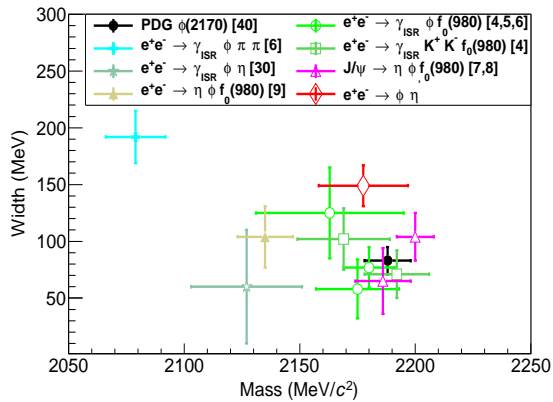}
\end{overpic}
\end{center}
\caption{The parameters of the $\phi(2170)$ state obtained from different processes and the resonance in the $e^{+}e^{-} \to \phi \eta^{\prime}$ process. }
\label{ComparisonOfParameters}
\end{figure}


The systematic uncertainties of the resonance parameters are mainly due to the signal model. To assess this systematic uncertainty, a modified BW function with mass-dependent width is used for the fit, resulting in differences of 19.5 MeV/$\it{c}^{\mathrm{2}}$ and 8.9 MeV for the mass and width, respectively. The dependence on the c.m.~energy determination and the fit procedure were also investigated and found to be negligible. The uncertainties (statistical and systematic) of the measured Born cross sections have been considered in the fit. Figure~\ref{ComparisonOfParameters} shows the comparison of the parameters of the $\phi(2170)$ state measured by experiments via various processes.

\section{Summary and Discussion}

 In summary, we present a precise measurement of the cross section line shape for $e^{+}e^{-} \to \phi \eta^{\prime}$ based on data samples collected with the BESIII detector at the BEPCII collider at 20 different c.m.~energies from 2.050 to 3.080 GeV. A clear structure is observed in the line shape of the measured Born cross sections. Assuming that the $\phi \eta^{\prime}$ comes from a single resonance, we determine the mass and width of this resonance to be (2177.5 $\pm$ 5.1 (stat) $\pm$ 18.6 (syst)) MeV/$\it{c}^{\mathrm{2}}$ and (149.0 $\pm$ 15.6 (stat) $\pm$ 8.9 (syst)) MeV, respectively. Here, the first uncertainties are statistical and the second ones are systematic. The statistical significance of the resonance is estimated to be larger than 10 $\sigma$, including systematic uncertainties. The $J^{PC}$ of the resonance should be  $1^{--}$ since it is produced in formation via $e^{+}e^{-}$ collisions. The mass of the resonance is compatible with the $\phi(2170)$. 
 
For the $2^{3}D_{1}$ $s\bar{s}$ excited state and the $\Lambda \bar{\Lambda}$ bound states in the molecular scenario, the decay mode of $\phi(2170) \to K^{+}K^{-}$ is favored. However, the parameters of the resonance extracted from the cross sections of $e^{+}e^{-} \to K^{+}K^{-}$ deviates from almost all individual measurements~\cite{ResoanceBESIII2019}. Thus, $\phi(2170)$ as a $2^{3}D_{1}$ $s\bar{s}$ quarkonium is disfavored. The width of the $3^{3}S_{1}$ $s\bar{s}$ is predicted to be about 380 MeV~\cite{Y2175SSbar2}, hence it cannot be identified with~$\phi(2170)$. Assuming that the observed resonance is the $\phi(2170)$, the measured $\mathcal{B}^{R}_{\phi \eta} \Gamma^{R}_{ee}$ = (1.7 $\pm$ 0.7 (stat) $\pm$ 1.3 (syst)) eV by {\it BABAR}~\cite{PRDPhiEta2008} is smaller than that of $\phi \eta^{\prime}$ mode. The ratio ($\mathcal{B}^{R}_{\phi \eta}\Gamma^{R}_{ee})/({\mathcal{B}^{R}_{\phi \eta^{\prime}}\Gamma^{R}_{ee}})$ is estimated to be 0.23 $\pm$ 0.10 (stat) $\pm$ 0.18 (syst). This is smaller than the prediction of the $s\bar{s}g$ hybrid models by several orders of magnitude~\cite{Y2175hybrid, Y2175hybrid3} and casts severe doubt on the validity of these models. We can not draw a conclusion about the other interpretations based on the current experimental results and need to perform further measurements in the future.


The BESIII collaboration thanks the staff of BEPCII and the IHEP computing center and the supercomputing center of USTC for their strong support. This work is supported in part by National Key Basic Research Program of China under Contract No.~2015CB856700; National Natural Science Foundation of China (NSFC) under Contracts No.~11625523, No.~11635010, No.~11735014, No.~11425524, No.~11335008, No.~11375170, No.~11475164, No.~11475169, No.~11605196, No.~11605198, No.~11705192; National Natural Science Foundation of China (NSFC) under Contract No.~11835012; the Chinese Academy of Sciences (CAS) Large-Scale Scientific Facility Program; Joint Large-Scale Scientific Facility Funds of the NSFC and CAS under Contracts No.~U1532257, No.~U1532258, No.~U1732263, No.~U1832207, No.~U1532102, No.~U1832103; CAS Key Research Program of Frontier Sciences under Contracts No.~QYZDJ-SSW-SLH003, No.~QYZDJ-SSW-SLH040; 100 Talents Program of CAS; INPAC and Shanghai Key Laboratory for Particle Physics and Cosmology; German Research Foundation DFG under Contract No.~Collaborative Research Center CRC 1044, FOR 2359; Istituto Nazionale di Fisica Nucleare, Italy; Koninklijke Nederlandse Akademie van Wetenschappen (KNAW) under Contract No.~530-4CDP03; Ministry of Development of Turkey under Contract No.~DPT2006K-120470; National Science and Technology fund; The Knut and Alice Wallenberg Foundation (Sweden) under Contract No.~2016.0157; The Swedish Research Council; U. S. Department of Energy under Contracts No.~DE-FG02-05ER41374, No.~DE-SC-0010118, No.~DE-SC-0012069, No.~DE-SC-0010504; University of Groningen (RuG) and the Helmholtzzentrum fuer Schwerionenforschung GmbH (GSI), Darmstadt



\end{document}